\documentclass[floatfix,lengthcheck,showpacs,amssymb,amsmath,amsfonts,twocolumn]{aastex63}
\usepackage{CJK}
\usepackage{lineno}
\usepackage[utf8]{inputenc}
\usepackage{color}
\usepackage{graphicx}
\usepackage{float}
\usepackage{subfigure}
\usepackage{amsmath}
\usepackage{afterpage}
\usepackage{acronym}
\usepackage{multirow}
\usepackage{tikz}
\usetikzlibrary{arrows,shapes,trees,decorations.pathreplacing}
\usepackage{import}
\usepackage{svn-multi}
\usepackage{hyperref}
\usepackage{gensymb}
\usepackage{longtable}
\hypersetup{
    colorlinks=true,
    linkcolor=blue,
    filecolor=magenta,      
    urlcolor=cyan,
}

\newcommand{\msun}{\ensuremath{\mathrm{M}_{\odot}}}

\usepackage{xcolor,colortbl}
\definecolor{o1}{gray}{0.9}
\definecolor{o2}{gray}{0.8}
\definecolor{o3}{gray}{0.6}
\definecolor{Gray}{gray}{0.9}
\usepackage{longtable}

\begin{document}
\begin{CJK*}{UTF8}{gbsn}
\title[]{Pre-merger localization of compact-binary mergers with third generation observatories}

\correspondingauthor{Alexander H. Nitz}
\email{alex.nitz@aei.mpg.de}

\author[0000-0002-1850-4587]{Alexander H. Nitz}
\affil{Max-Planck-Institut f{\"u}r Gravitationsphysik (Albert-Einstein-Institut), D-30167 Hannover, Germany}
\affil{Leibniz Universit{\"a}t Hannover, D-30167 Hannover, Germany}
\author[0000-0001-5078-9044]{Tito Dal Canton}
\affil{Universit{\'e} Paris-Saclay, CNRS/IN2P3, IJCLab, 91405 Orsay, France}


\begin{abstract}
We present the prospects for the pre-merger detection and localization of binary neutron star mergers with third generation gravitational-wave observatories. We consider a wide variety of gravitational-wave networks which may be operating in the 2030's and beyond; these networks include up to two Cosmic Explorer sites, the Einstein Telescope, and continued observation with the existing second generation ground-based detectors. For a fiducial local merger rate of 300 Gpc$^{-3}$yr$^{-1}$, we find that the Einstein Telescope on its own is able to detect 6 and 2 sources per year at 5 and 30 minutes before merger, respectively, while providing a localization of $<10~\textrm{deg}^2$. A single Cosmic Explorer would detect but be unable to localize sources on its own. A two-detector Cosmic Explorer network, however, would detect 22 and 0.4 mergers per year using the same criteria. A full three-detector network with the operation of dual Cosmic Explorers and the Einstein Telescope would allow for $<1~\textrm{deg}^2$ source localization at 5 minutes before merger for $\sim7$ sources per year. Given the dramatic increase in localization and detection capabilities, third generation observatories will enable the regular observation of the prompt emission of mergers by a broad array of observatories including gamma-ray, x-ray, and optical telescopes. Moreover, sub-degree localizations minutes before merger, combined with narrow-field-of-view high-energy telescopes, could strongly constrain the high-energy pre-merger emission models proposed in the last decade.
\end{abstract}

\section{Introduction}

Gravitational-wave (GW) astronomy has rapidly matured since its dawn in 2015~\citep{TheLIGOScientific:2016qqj,Abbott:2016blz}. There are now dozens of observed binary black hole mergers~\citep{Abbott:2020niy,LIGOScientific:2018mvr,Nitz:2021uxj} along with the two binary neutron star mergers, GW170817~\citep{TheLIGOScientific:2017qsa} and GW190425~\citep{Abbott:2020uma}. Most recently, two neutron star -- black hole mergers have been reported~\citep{LIGOScientific:2021qlt}. As GW observation becomes a standard component of astronomy, there remain exciting opportunities for coordinated multi-messenger observation. The first significant example was GW170817 which was observed simultaneously in gamma-rays~\citep{Goldstein:2017mmi,Monitor:2017mdv,Savchenko:2017ffs,LIGOScientific:2017zic} with numerous x-ray, optical, and radio observations following afterward~\citep{GBM:2017lvd}.

One of the expected outcomes of GW astronomy is the early (pre-merger) detection of an imminent stellar-mass compact binary merger \citep{Cannon2011Early, Sachdev:2020lfd, Nitz:2020vym, Magee:2021xdx}. Such a pre-merger detection, especially if associated with a sufficiently precise sky localization of the binary, would allow sensitive, narrow-field electromagnetic telescopes to observe the merger and the entire development of any electromagnetically-bright postmerger process, like a kilonova in the case of a binary neutron star (BNS) system \citep{Metzger:2019zeh}. Pre-merger detections of BNS systems can already be achieved with existing second-generation (2G) ground-based interferometers; these include the twin Advanced LIGO observatories \citep{AdvLIGO}, Advanced Virgo~\citep{AdvVirgo}, KAGRA \citep{KAGRA2020}, and the under-construction LIGO India~\citep{Unnikrishnan:2013qwa,Saleem:2021iwi}. The advance warning time, however, depends critically on the sensitivity of the detectors in the frequency band below $\sim 20$ Hz. For 2G observatories, even with substantial upgrades, this time is expected to rarely exceed 10 minutes, especially when requirements on the sky localization are imposed \citep{Nitz:2020vym}.

However, third-generation (3G) detectors like Cosmic Explorer (CE; \cite{cewhitepaper}) and Einstein Telescope (ET; \cite{et}) are expected to significantly increase the low-frequency sensitivity compared to 2G observatories, and therefore raise the possibility of routine pre-merger detections and precise localizations of compact binaries. Various studies have already investigated the expected performance of 3G detectors \citep{Mills:2017urp, Chan:2018csa, Grimm:2020ivq, Tsutsui:2020bem}. One common feature of previous studies which examined 3G pre-merger localization is the use of the Fisher matrix approach as a computationally-inexpensive way to estimate the uncertainties in the sky localization.

In this paper, we present a systematic exploration of the pre-merger detection capabilities of a comprehensive set of future networks, composed of combinations of 2G and 3G detectors. Instead of relying on the Fisher matrix approach, we apply full Bayesian parameter estimation to robustly characterize the posterior distribution of the binary's spatial location. In addition, we use the latest estimates for the detectors' sensitivities (see Sec.~\ref{sec:det} for further details). We show detection rates as a function of the advance warning time for the various considered networks, for different requirements on the precision of the sky localization, and for different requirements on the luminosity distance to the source.

\begin{table*}
    \caption{The location, noise curve, and low frequency cutoff $f_{\textrm{low}}$~used for the detector configurations we consider in this paper. Note that several observatories are listed twice as we consider different detector configurations at the same site. The locations are rounded. For sites which are not yet constructed, the locations are fiducial, however, we expect the distances between the detectors to be representative; we do not expect the localization capabilities to significantly differ from the final site selection.}
    \label{table:detectors}
\begin{center}
\begin{tabular}{llllcc}
Abbreviation & Observatory  & $f_{\textrm{low}}$ &  Noise Curve & Latitude & Longitude \\ \hline
H & LIGO Hanford & 7 & Voyager & 46.5 & -119.4 \\
L & LIGO Livingston & 7 & Voyager & 30.6 & -90.8 \\
I & LIGO India & 7 & Voyager & 14.2 & 76.4 \\
V & Virgo & 11 & O5 High &  43.6 & 10.5 \\
K & KAGRA & 11 & Design &  36.4 & 137.3 \\
$C_1^U$ & Cosmic Explorer USA & 5.2 & CE1 & 40.8 & -113.8   \\
$C_1^A$ & Cosmic Explorer Australia & 5.2 & CE1 & -31.5 & 118.0  \\
$C_2^U$ & Cosmic Explorer USA & 5.2 & CE2 & 40.8 & -113.8    \\
$C_2^A$ & Cosmic Explorer Australia & 5.2 & CE2 & -31.5 & 118.0   \\
$E$ & Einstein Telescope & 2 & ET-D Design & 43.6 & 10.5  \\
$E_5$ & Einstein Telescope & 5 & ET-D Design& 43.6 & 10.5  \\

\end{tabular}
\end{center}
\end{table*}

\section{Detector Networks}
\label{sec:det}

In this work, we consider various detector networks that include combinations of both 2G and 3G observatories. Table~\ref{table:detectors} lists each detector configuration that we consider, along with its location, noise curve, lower frequency cutoff, and named abbreviation (for comparison to later figures). We assume all detectors have 100\% duty cycle.

For the 2G detectors we have chosen noise curves which represent recent knowledge about possible future sensitivity. For the three LIGO detectors (Hanford, Livingston, India) this represents their "Voyager" configuration which corresponds to significant in-site upgrades which may be constructed towards the end of this decade or into the early 2030's~\citep{voyager}. For Virgo and KAGRA, we use the optimistic O5 sensitivity curves~\citep{o5noise,Abbott:2020qfu}.

Cosmic Explorer is proposed to be constructed in two phases with increasing design complexity and observing sensitivity, known as CE1 and CE2~\citep{Hall:2020dps,CENoise}. In addition to these two sensitivity targets, we consider the possibility of a second Cosmic Explorer site in Australia. For the US and Australian sites, we use the fiducial site locations from ~\cite{Hall:2019xmm}. There is also a proposal for another GW detector, NEMO, to be constructed in Australia~\citep{Ackley:2020atn}. We have not considered NEMO in this work due to its focused goal of high-frequency observation, however, NEMO may act as a consumer of pre-merger alerts to enable optimal observing of the BNS post-merger signals. The Einstein Telescope has the best low-frequency sensitivity of the 3G proposals~\citep{Hild:2010id,ETnoise}. In this paper we consider both its full design sensitivity and also the case where only the sensitivity above 5 Hz is reached.

\section{Source Population and Parameter Estimation}

We consider a fiducial population of 1.4-1.4~\msun~ (source-frame) BNS mergers which is distributed isotropically in orientation and sky location. We assume a fiducial, local merger rate of $300\, \textrm{Gpc}^{-3}\textrm{yr}^{-1}$, which is consistent with the observational constraints from current GW searches~\citep{Abbott:2020niy}. We assume the rate of mergers evolves with redshift following the fitting formula for the star formation rate density of~\cite{Madau_2014}. As suggested in \cite{Nitz:2020vym}, scaling relations between the time and rate can be used to translate our results for alternate mass choices or merger rate assumptions. 

We use PyCBC Inference~\citep{Biwer:2018osg,pycbc-github} and a heterodyne likelihood model~\citep{Cornish:2010kf,Finstad:2020sok,zackay2018relative} to estimate the parameters of our simulated population of GW sources. Sampling of the parameter space is done using the Dynesty package which implements a nested sampling algorithm~\citep{speagle:2019}. Because of the low frequency capabilities of third generation detectors (perhaps down to $\sim 2$ Hz with ET), we account for time variation in the antenna response of the signal due to the earth's rotation in our likelihood function. We assume standard isotropic priors for the sky location and orientation angles of each source. All signals are placed into simulated detector noise that is assumed to be stationary Gaussian noise and lacks non-Gaussian noise transients. Investigation of these choices is needed to produce robust analyses in the future, as long-duration signals may intersect numerous transient noise sources~\citep{Davis:2021ecd} or other signals~\citep{Samajdar:2021egv, Pizzati:2021gzd, Himemoto:2021ukb}. However, we do not expect possible noise or signal mitigation measures to significantly impact the statements here.

We use the TaylorF2 waveform approximant to model the GW signal. This model includes
only the dominant mode of the GW signal, is accurate to 3.5 post-Newtonian order~\citep{Sathyaprakash:1991mt,Droz:1999qx,Blanchet:2002av,Faye:2012we}, and neglects orbital precession due to misaligned spins. This model is appropriate for signals where the merger is excluded from the analysis, either implicitly due to the signal-to-noise at high frequencies, or explicitly due to imposing a cutoff. We estimate the localization properties by truncating our analysis at fixed times before merger (we consider 1-30 minutes before merger here) and performing the parameter estimation with a tapered and truncated signal model. We have confirmed that this procedure produces self-consistent posteriors by performing a percentile-percentile test.

We note that for high-mass-ratio sources, such as from neutron star -- black hole mergers, neglecting sub-dominant modes, as we do here, may lead to overly conservative estimates of the localization capabilities~\citep{Kapadia:2020kss}. A similar caveat applies to orbital precession as well~\citep{Tsutsui:2020bem}. For comparable-mass sources with small spins, however, we expect our predictions to apply. Exploration of the pre-merger localization capabilites of 3G detectors for sources with non-negligible signal-to-noise in the higher-order modes is left to future work.

We simulate a population of $O(10^5)$ GW signals and consider only those signals which have a coherent SNR $> 12$ to be detected for purposes of this analysis. This is consistent with the choices in ~\cite{Magee:2021xdx,Nitz:2020vym}. We consider this a conservative choice, which allows for increased noise in future detector networks. We note that for well-localized sources, this constraint is not limiting. The median coherent SNR is greater than $\sim20$ for all detector networks considered and advance warning time from 1-30 minutes if we restrict to sources that can be localized to $<100~\textrm{deg}^2$; the median SNR is higher if we impose tighter localization requirements.

While the pre-merger detection of sources which can be well-localized should be easily achieved, there remain significant challenges in overall coordination between the GW and follow-up observatories~\citep{Magee:2021xdx} and importantly in developing rapid localization capabilities. Depending on the network configuration, our analysis running on a single core requires between several hours to days to determine the parameters and localize a single source. The method used here would be insufficient to provide early-warning localizations; an extension of the rapid techniques developed in \cite{BAYESTAR} would be needed.

\section{Detection and Localization Capabilities}

\begin{figure*}
  \centering
  \includegraphics[width=\columnwidth]{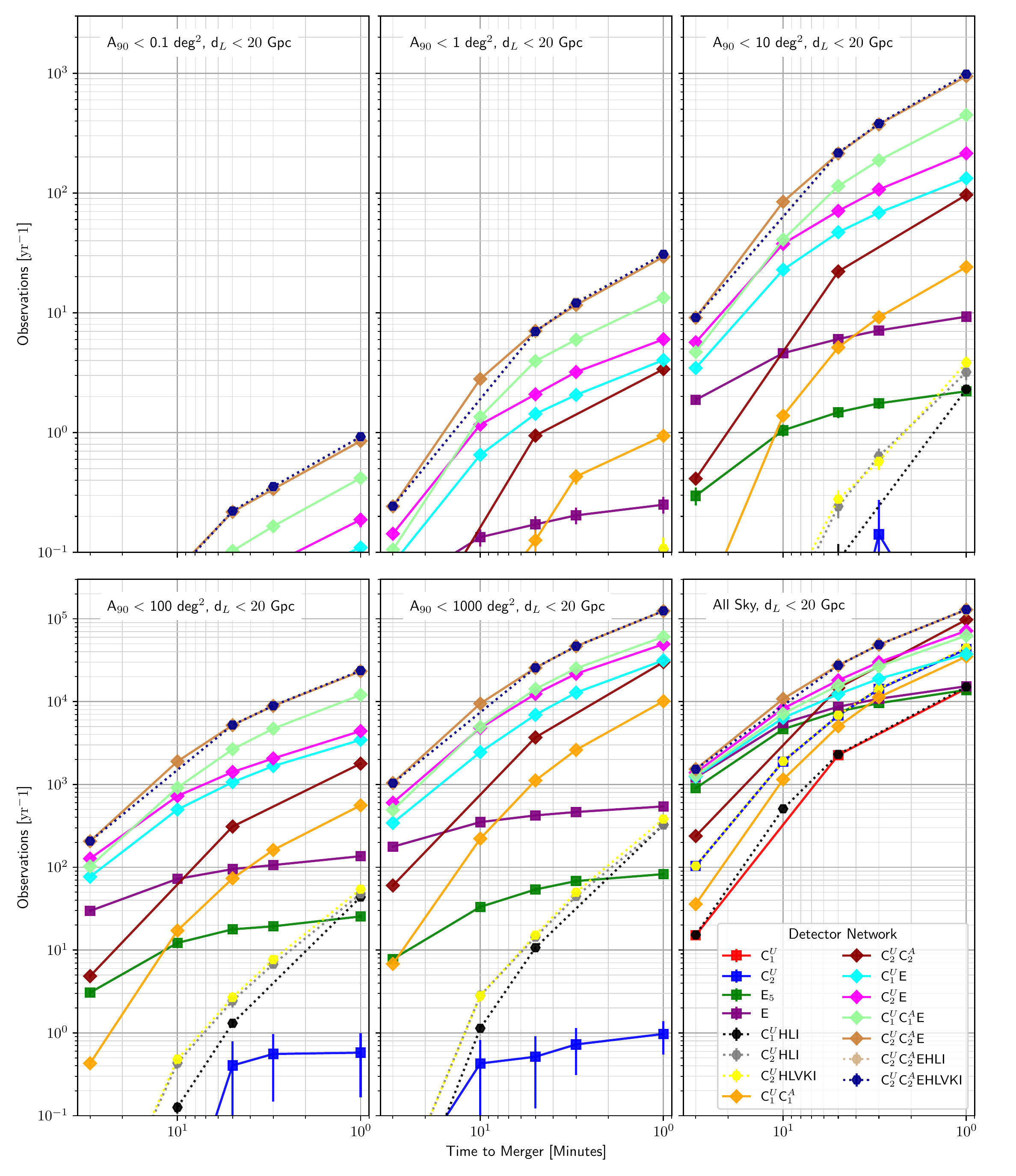}
\caption{The number of sources detected as a function of time before merger that are within a luminosity distance $d_L$ of 20 Gpc ($z\sim2.4$) for a fiducial population of 1.4-1.4~\msun~BNS mergers whose rate tracks the star formation and has a local density of 300 Gpc$^{-3}$yr$^{-1}$. Results are shown for various constraints on the localization area ($A_{90}$) that is possible to achieve using only data that is available up to that point in time. Localization areas are reported using a $90\%$ credible region. Some networks do not appear in all panels due to failing to detect mergers that satisfy the constraints within the plotted merger rate limits.}
\label{fig:cap20000}
\end{figure*}

\begin{figure*}
  \centering
  \includegraphics[width=\columnwidth]{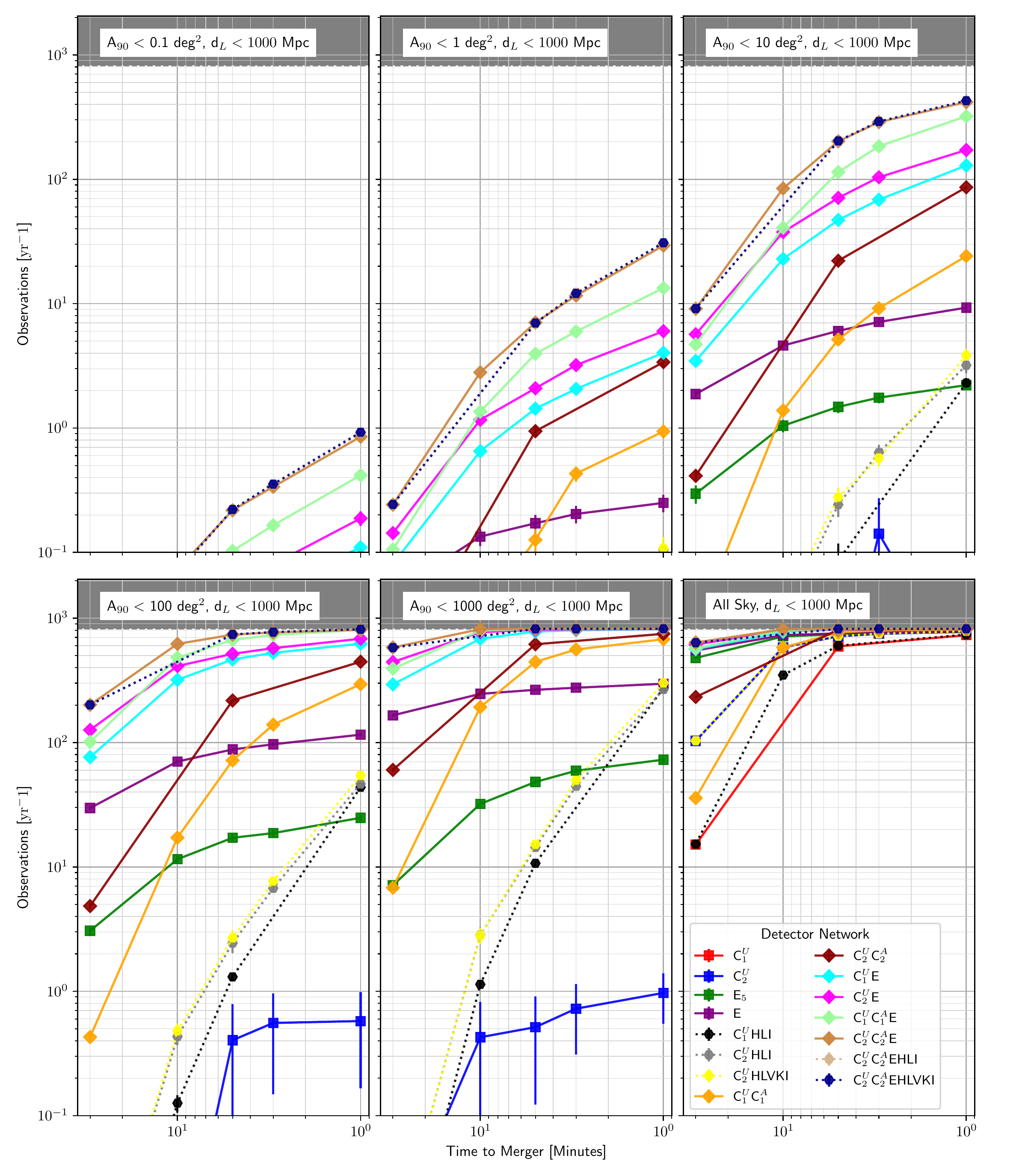}
\caption{The number of sources detected as a function of time before merger that are within a luminosity distance $d_L$ of 1000 Mpc ($z\sim0.2$) for a fiducial population of 1.4-1.4~\msun~BNS mergers whose rate tracks the star formation and has a local density of 300 Gpc$^{-3}$yr$^{-1}$. Results are shown for various constraints on the localization area ($A_{90}$) that is possible to achieve using only data that is available up to that point in time. Localization areas are reported using a $90\%$ credible region. Some networks do not appear in all panels due to failing to detect mergers that satisfy the constraints within the plotted merger rate limits. The merger rates excluded by the distance constraint are shaded in gray. For a three-detector 3G network, we can see that nearly all sources are detected and localized to $<100~\textrm{deg}^2$ at 5 minutes before merger.}
\label{fig:cap1000}
\end{figure*}

\begin{figure*}
  \centering
  \includegraphics[width=\columnwidth]{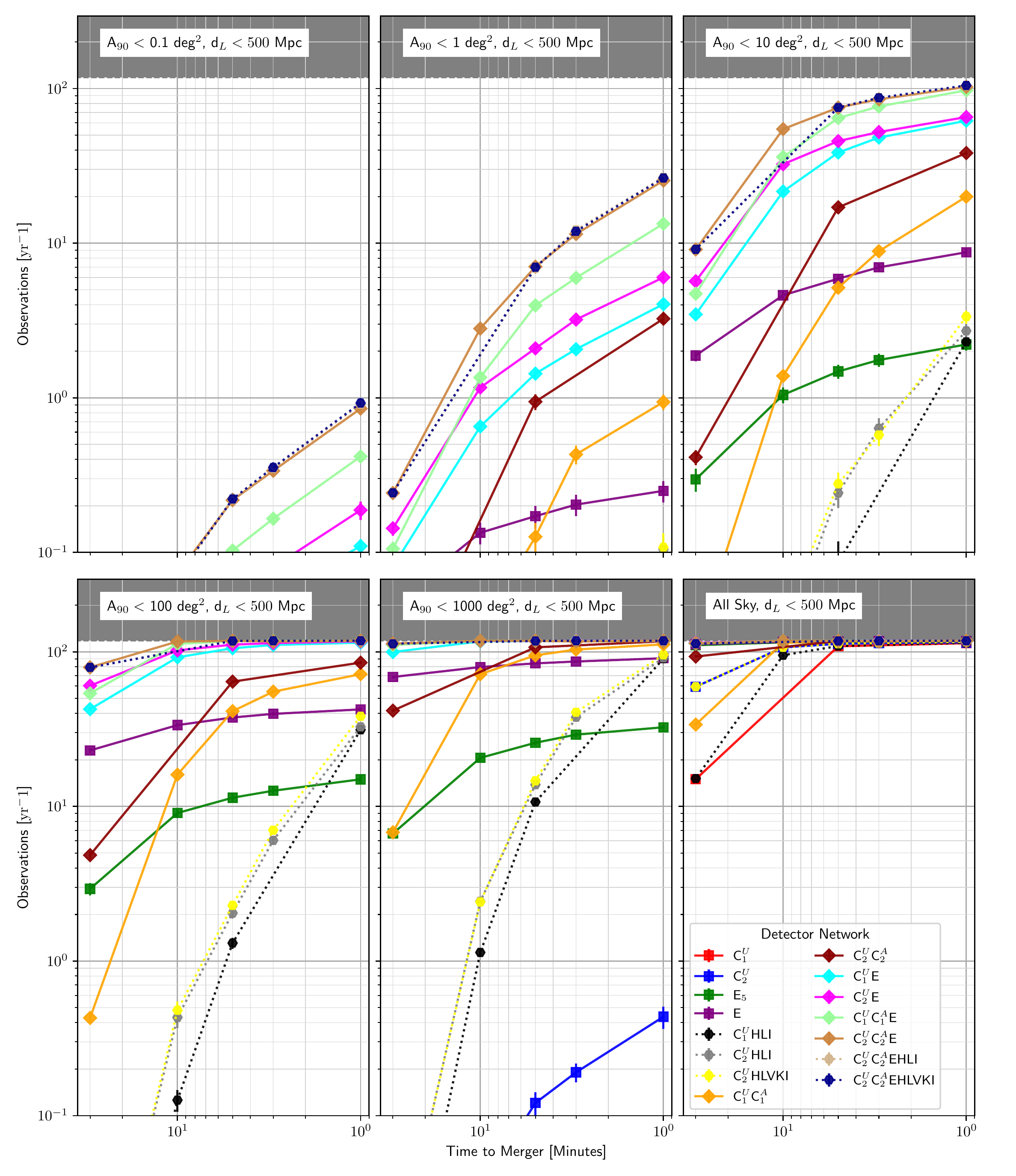}
\caption{The number of sources detected as a function of time before merger that are within a luminosity distance $d_L$ of 500 Mpc ($z\sim0.1$) for a fiducial population of 1.4-1.4~\msun~BNS mergers whose rate tracks the star formation and has a local density of 300 Gpc$^{-3}$yr$^{-1}$. Results are shown for various constraints on the localization area ($A_{90}$) that is possible to achieve using only data that is available up to that point in time. Localization areas are reported using a $90\%$ credible region. Some networks do not appear in all panels due to failing to detect mergers that satisfy the constraints within the plotted merger rate limits. The merger rates excluded by the distance constraint are shaded in gray. For a three-detector 3G network, we can see that the majority of sources within this distance are detected and localized to $<10~\textrm{deg}^2$ at 5 minutes before merger.}
\label{fig:cap500}
\end{figure*}

\begin{figure*}
  \centering
  \includegraphics[width=\columnwidth]{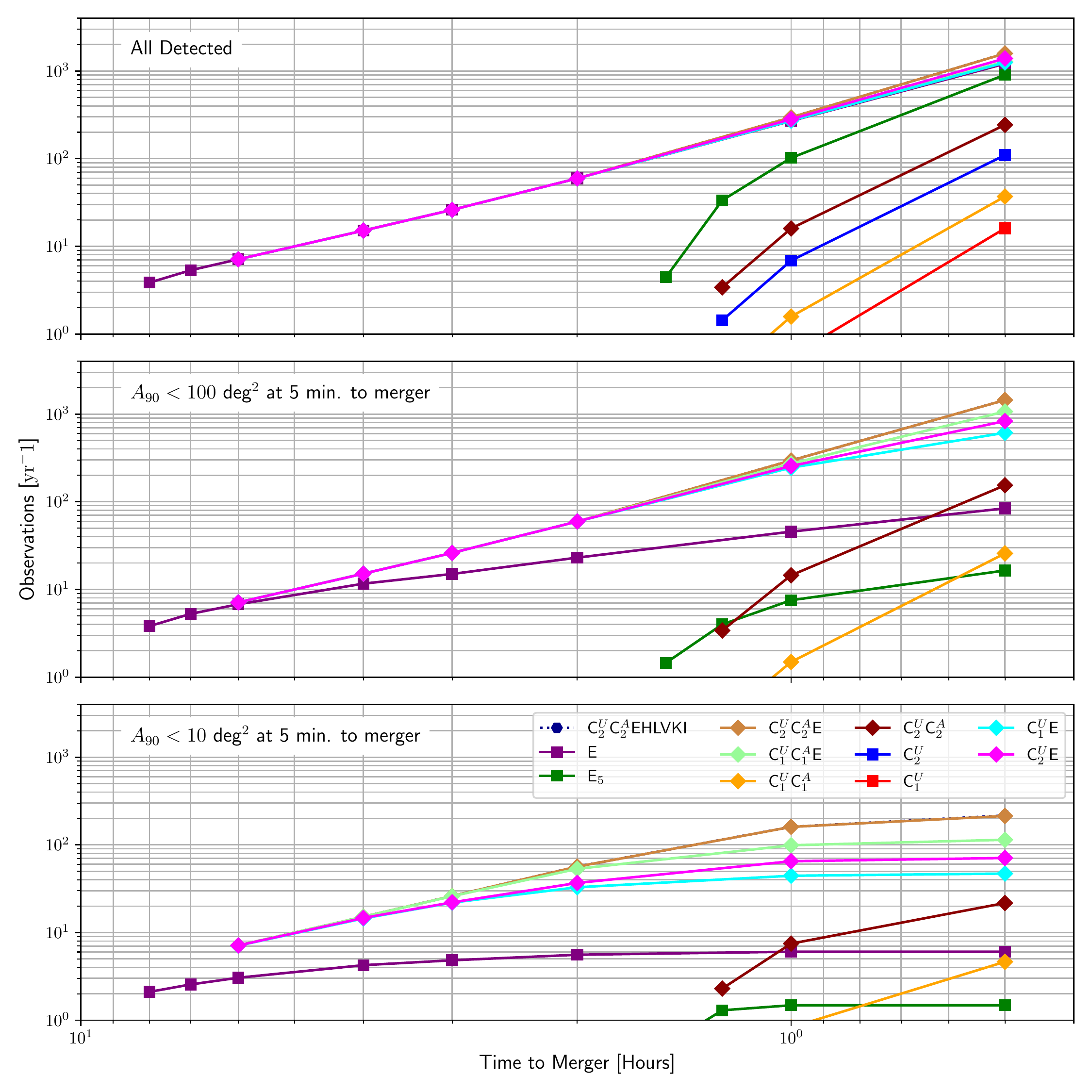}
\caption{The number of sources detected as a function of time before merger for our fiducial population of 1.4-1.4~\msun~BNS mergers whose rate tracks the star formation and has a local density of 300 Gpc$^{-3}$yr$^{-1}$. Results are shown for various constraints on the localization area ($A_{90}$) these detected sources later achieve at 5 minutes prior to merger. Multi-hour advance warning of mergers is possible with the Einstein Telescope. Advance warning of sources which will eventually have a precise localization may provide time for other observatories to be brought into observing mode, reschedule observations, or allow for manual human intervention as needed.}
\label{fig:long}
\end{figure*}

The number of sources per year which can be observed and localized as a function of time before merger are shown in Figs~\ref{fig:cap20000}-\ref{fig:cap500} for each detector network. We consider various constraints on the localization area, as well as limits on the luminosity distance at 20 Gpc (Fig.~\ref{fig:cap20000}), 1 Gpc (Fig.~\ref{fig:cap1000}), and 500 Mpc (Fig.~\ref{fig:cap500}). These figures focus on the localization abilities within 30 minutes of merger; as shown in Fig.~\ref{fig:long}, however, it is possible to detect sources hours before merger, especially those that will at late-times be well-localized. Detection of a source several hours before merger may allow for observatories, both GW and non-GW, to ensure readiness at the merger time. For example, there may be sufficient time to cancel previously scheduled commissioning or maintenance activities which would have otherwise rendered an observatory offline.

\subsection{Single Cosmic Explorer}

We have considered both the CE1 and CE2 designs. Both configurations will be able to give 5 minutes advance warning of $> 1000$ mergers per year. However, we can see that if Cosmic Explorer operates without additional observatories, the ability to localize these sources will be limited; in the best case CE2 on its own could localize a single source to a $<1000~\textrm{deg}^2$ area per year. If the existing second-generation sites are able to continue observation, however, then the combined network including a single CE1 or CE2 is able localize O(1) sources per year with 5 minutes of warning and $<100~\textrm{deg}^2$ localization. It would be useful for this purpose to maintain operation of second-generation observatories.

\subsection{Einstein Telescope}

The Einstein Telescope has a unique triangular design with effectively three independent interferometers at the same location~\citep{Punturo:2010zz}. Of the third-generation designs, it also has the most aggressive targets for low-frequency sensitivity (down to $\sim 2 $ Hz). This means that neutron star signals can be observed for many hours and effects such as the rotation of the Earth can become important. The expectation is that it may be a significant challenge to achieve these low-frequency goals~\citep{Beker:2014ata,Badaracco:2021ory}. If they are achieved, however, we find that the construction of Einstein Telescope would enable $\sim 6$ or $94$ sources to be localized 5 minutes before merger to an area of $<10$ or 100~$\textrm{deg}^2$ per year, respectively. If only the sensitivity above 5 Hz is achieved, this rate is reduced by a factor of $\sim4-5$. 

\subsection{Multiple 3G Observatories}

If multiple third-generation observatories are built, the localization capabilities significantly increase. A combined USA-Australian network of two Cosmic Explorers is able to localize 1, 22, or 310 sources per year at 5 minutes before merger and $<$1, 10, or 100 deg$^2$ localization area, respectively. The dual Cosmic Explorer network is able to outperform networks with the Einstein Telescope as the lone 3G observatory due to available timing information and overall higher signal SNR at late times. This network underperforms ET on its own at early times due to its more limited low-frequency sensitivity. The localization capabilities increase to 7, 210, and 5200 sources per year, respectively, with 5 minutes of advance warning if the Einstein Telescope is included as a part of a three-detector 3G network. This three-detector 3G network would also provide a significant number of precise localizations at 30 minutes prior to merger, with 0.3, 9, and 200 sources per year when requiring a $<$1, 10, or 100 deg$^2$ localization area, respectively. This network is also the only capable of providing regular sub-degree localizations before merger; a single source per year could be detected and localized one minute before merger with sky area $<0.1~ \textrm{deg}^2$.

\subsection{Considering Alternate Source Masses}

Due to the uncertain distribution of masses for binary neutron star sources, we have focused on a fiducial 1.4-1.4~\msun~BNS merger population. However, as noted in ~\cite{Nitz:2020naa}, these localization results can be straightforwardly rescaled to apply to sources with other masses or other models of the merger rate. This rescaling holds where the target source can still be modelled as a post-Newtonian inspiral at a chosen time before merger. To map our results onto another reference source, one would scale the time axis and the rate axis using the relations

\begin{equation}
    T_{m_1,m_2} = T_{1.4-1.4} \frac{2.8\msun}{m_1 + m_2},
\end{equation}
\begin{equation}
    R_{m_1,m_2} = R_{1.4-1.4} \frac{(m_1 m_2)^{3/2}}{(m_1 + m_2)^{1/2}} \frac{2.8^{1/2}}{1.4^3}.
\end{equation}

where $m_{1,2}$ are the component masses of the target desired source and $T_{1.4-1.4}$ and $R_{1.4-1.4}$ are the time before merger and rate, respectively, for the fiducial 1.4-1.4~\msun merger in our results.

By extension, these scaling relations could also be used to determine the expected detection rates and localization properties for extended mass distributions. For population models with the same overall merger rate and a distribution whose typical mass is similar to our fiducial, the results would not be substantially changed from those shown here. However, if the population were to shift the majority of sources to higher (lower) masses this would tend to increase (decrease) the expected number of well-localized detections several minutes before merger, while having the opposite effect very close to merger.

\section{Conclusions}

We provide guidance on the advance warning capabilities provided by the GW observatory networks of the 2030-40's to help enable  multi-messenger observation of the prompt post-merger emission of BNS mergers. We demonstrate the precise capabilities of 3G GW networks by simulating a population of BNS mergers and using Bayesian parameter estimation to determine their localization estimates at 1-30 minutes before merger. 

We show that advance warning with accurate localizations will become commonplace. With the possibility of O(1), O(10), or O(1000) detected sources per year with localizations $<$ 1, 10, or 100 deg$^2$, respectively, and 5 minutes of advance warning or more, it should be within the capabilities of numerous observatories to regularly repoint and directly observe the prompt emission.
Monitoring the last few minutes of several BNS inspirals with sensitive, narrow-field high-energy telescopes would also test the existence of pre-merger electromagnetic transients, a hypothesis proposed various times in the last decade but currently unsettled \citep[e.g.][]{BurnsNS}.
The sensitivity of telescope and observatory follow-up partners is a potential area of concern that should be factored into future proposals; $\sim 90\%$ of the sources with localization area of $<100~\textrm{deg}^2$ at 5 minutes before merger are found at distances $>1~\textrm{Gpc}$. The most well-localized sources, however, will tend to be close-by; the vast majority of sources localized to $<1~\textrm{deg}^2$ will lie within only a few hundred megaparsecs. 

To aid in matching our results with specific follow-up observatory capabilities and for planning of future proposals, we make the data associated with the simulations along with the scripts to reproduce the figures in this paper available at~\url{https://github.com/gwastro/gw-3g-merger-forecasting}.
\acknowledgments

 We are grateful to the computing team from AEI Hannover for their significant technical support with special thanks to Carsten Aulbert and Henning Fehrmann.

 Software: Numpy \citep{vanderWalt:2011bqk}, Scipy \citep{2020SciPy-NMeth}, Astropy \citep{Robitaille:2013mpa}, Matplotlib \citep{matplotlib}, Dynesty~\citep{speagle:2019}, PyCBC \citep{pycbc-github}.
 
\bibliography{references}

\end{CJK*}
\end{document}